\documentclass[a4paper,11pt]{article}
\pdfoutput=1 
\usepackage{jcappub} 
\usepackage{xcolor}
\usepackage{verbatim}
\usepackage{amsmath}
\usepackage{amsthm}
\usepackage{chngcntr}

\usepackage{multirow}

\title{\boldmath Sub-GeV dark matter in neutron stars: halo morphologies and their suppression by vacuum–like pressure}

\author[1]{Loreany F. Araújo,\href{https://orcid.org/0000-0003-1749-8354}{\includegraphics[scale=0.1]{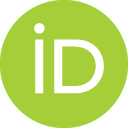}}}
\affiliation[1]{Departamento de Astronomia, Universidade de S\~{a}o Paulo \\ Rua do Mat\~{a}o, 1226 - 05508-900, S\~ao Paulo, SP, Brazil}
\author[2]{Germ\'{a}n Lugones,\href{https://orcid.org/0000-0002-2978-8079}{\includegraphics[scale=0.1]{ORCIDiD_icon128x128.png}}}
\affiliation[2]{Universidade Federal do ABC, Centro de Ci\^encias Naturais e Humanas \\ Avenida dos Estados 5001- Bang\'u, CEP 09210-580, Santo Andr\'e, SP, Brazil}
\author[1]{and José Ademir S. Lima\href{https://orcid.org/0000-0001-5426-3197}{\includegraphics[scale=0.1]{ORCIDiD_icon128x128.png}}}

\emailAdd{loreanyfa@usp.br, german.lugones@ufabc.edu.br, jas.lima@iag.usp.br}

\abstract{We investigate neutron stars that contain a unified dark sector composed of cold, degenerate fermionic dark matter and a vacuum–like dark–energy component. Within a general–relativistic two–fluid framework that allows a covariantly conserved, gradient–driven energy exchange between baryons and the dark sector, we quantify how dark microphysics reshapes global structure when the total gravitational radius need not coincide with the luminous baryonic radius. Using a state-of-the-art baryonic equation of state, we explore the halo–forming mass range for fermionic dark matter with particle masses of 400 MeV and 1 GeV, and we characterize sequences by the difference between the total and luminous radii and by the fractional difference between the total and baryonic masses. We confirm established trends: lighter fermions typically support low-density halos that increase the total radius by several kilometers at nearly fixed mass, whereas masses near 1 GeV tend to shrink halos and make the two radii appreciably closer. Our central new result is that a percent-level vacuum–like admixture markedly reduces halo formation, shrinking the radius difference from several kilometers to sub–kilometer scales and the fractional mass difference to $\lesssim 1\%$. Combined gravitational-wave and X-ray observations offer a practical route to bound the halo size and the allowed vacuum–like fraction.}

\begin{document}
\maketitle
\flushbottom

\section{Introduction}
\label{sec:intro}

The microscopic nature of dark matter (DM) and dark energy (DE) remains a central open problem in fundamental physics and cosmology. While a broad array of cosmological and galactic observations tightly constrains their large–scale gravitational effects, any non–gravitational couplings to the Standard Model are poorly known \citep{Bertone:2004pz,Schumann:2019eaa,Bertone:2018xtm}. Neutron stars (NSs) provide a complementary arena to probe such interactions: their deep gravitational potentials, supranuclear densities, and long lifetimes can amplify even feeble dark couplings, yielding cumulative imprints on structure and dynamics that are, in principle, accessible to multimessenger observations.

A rich set of mechanisms has been proposed to test DM with NSs. Capture via scattering on stellar constituents can produce a bound DM population whose annihilation or kinetic heating modifies the heat budget and cooling history of otherwise cold, old NSs \citep{Guver:2014JCAP,Goldman:1989nd,Kouvaris:2007ay,Bell:2024JCAP,Bramante:2024PhysRep}. Additional possibilities include dark or baryon–number–violating decays in the interior \citep{Baym:2018wtr}, compact–object coalescences involving mirror or DM–admixed NSs with distinctive gravitational–wave signatures \citep{Hippert:2023Dark}, and scenarios in which annihilation products (e.g., long–lived mediators) escape the star and generate electromagnetic or neutrino emission outside the surface, reshaping the expected multimessenger phenomenology \citep{Leane:2017vag,Acevedo:2025JCAP,Dasgupta:2020jxi}. Beyond thermal and high–energy messengers, dark components may also shift NS oscillation spectra (e.g., $f$–modes), opening a route for asteroseismology and future gravitational–wave constraints \citep{Shirke:2024fmode}.

At the level of global structure, the impact of a dark component is cleanly described in a general–relativistic two–fluid framework, in which the dark and baryonic sectors possess distinct equations of state and, in general, distinct radial profiles. This viewpoint naturally organizes equilibria into compact configurations with a dark–dominated core, partially overlapping “admixed” profiles, and extended halos in which the dark density predominates outside the baryonic surface—an organizing picture already apparent in early two–fluid analyses \citep{Leung:2011zz,Panotopoulos:2017idn,Ellis:2018bkr}. Two–fluid studies with \emph{fermionic} DM map these regimes across parameter space: sub–GeV fermions tend to support low–density halos that inflate the \emph{gravitational} radius with comparatively small mass shifts, whereas near–GeV particles largely suppress halo formation and drive the configuration toward a baryonic baseline \citep{Leung:2022wcf,Nelson:2018xtr,Scordino:2024ehe,Li:2012ii}. Related analyses with bosonic DM and repulsive self–interactions display analogous trends in radius and tidal response, further restricting the admissible dark fraction \citep{Karkevandi:2021ygv}. Crucially, when diffuse halos are present, the radius inferred from space-time observables (sensitive to the exterior metric) does not need to coincide with the photospheric radius tied to the luminous surface; this distinction is essential for interpreting X–ray pulse profiles, self–lensing, and tidal–deformability constraints \citep{Shawqi:2024jmk,Liu:2024swd,Miao:2022rqj}.

The last decade has delivered a qualitative leap in NS astrophysics: radio timing of massive pulsars, gravitational–wave constraints on tidal deformability from GW170817, and NICER pulse–profile inferences now anchor the equation of state (EOS) at supranuclear densities with unprecedented precision \citep{Abbott:2018exr, Miller:2021qha, Miller:2019cac, Riley:2021pdl, Riley:2019yda}. This multimessenger toolkit can be repurposed to test dark–sector effects. In halo–dominated configurations—where the total (gravitational) radius grows more than the mass at fixed $M$—the tidal deformability $\Lambda=\tfrac{2}{3}k_2 C^{-5}$ is generically amplified, tightening GW bounds on the allowed halo extent \citep{Nelson:2018xtr,Karkevandi:2021ygv}. In contrast, pulse–profile modeling constrains a photospheric radius tied to the luminous surface, reinforcing the need to treat $R$ and $R_{\rm bm}$ as distinct observables in any confrontation with data.  Relatedly, a recent Bayesian analysis of NICER mass–radius data for neutron stars admixed with fermionic asymmetric dark matter found that, at current observational precision, configurations with dark cores are effectively indistinguishable from purely baryonic stars, while still allowing constraints on combinations of the dark self-repulsion strength and particle mass.  Importantly, that study explicitly excludes halo configurations—precisely the regime explored here—because halos alter the pulse profile and its interpretation \citep{Rutherford:2024uix}.

Dark–energy–like sectors can likewise influence hydrostatic balance when modeled as a vacuum–like component ($w\simeq -1$) or as part of a unified dark sector. In such cases, negative pressure can soften the effective force balance in the outer layers and reduce the radii at fixed mass, thus modifying the tidal properties and the reading of multi-messenger constraints \citep{Araujo:2025tlv}. In our previous work \citep{Araujo:2025tlv}, we quantified these effects for fermionic DM with benchmark mass $m_\chi=10~\mathrm{GeV}$, a regime that does not produce extended halos. There we showed that introducing a vacuum–like fraction leads to a systematic contraction of the sequences and a reduction of the maximum mass, with the luminous and gravitational radii remaining closely aligned due to the absence of a diffuse dark envelope. We further demonstrated that allowing energy exchange between baryons and the dark sector via a covariantly conserved, gradient–driven source term steepens the pressure decline in the exterior layers when a vacuum–like component is present, reinforcing the net compacting trend.

The present paper is a natural continuation of that program, but focused on the \emph{halo–forming} mass range that was not explored in \citep{Araujo:2025tlv}. Throughout, we label the sectors by the subscripts `$\mathrm{bm}$', `$\chi$', and `$\mathrm{de}$' for baryonic matter, DM, and DE, respectively. We move to lighter DM fermions, $m_\chi=\{400~\mathrm{MeV},\,1~\mathrm{GeV}\}$, for which two–fluid equilibria can develop low–density halos that inflate the total (gravitational) radius $R$ relative to the luminous baryonic radius $R_{\rm bm}$. For clarity, we distinguish two radii and two masses: (i) $R_{\rm bm}$ is the location of the baryonic surface; the associated mass $M_{\rm bm}$ is the gravitating mass enclosed within $R_{\rm bm}$—including any DM/DE present inside—and is the quantity constrained by radiative observables from that surface; (ii) $R$ is the outer boundary of the configuration, defined by the outermost radius at which the total pressure $p_{\rm bm}(r)+p_\chi(r)+p_{\rm de}(r)$ vanishes, and $M$ is the total gravitational mass enclosed within $R$. We introduce data–facing diagnostics—radius and mass excesses $\Delta R\equiv R-R_{\rm bm}$ and $\Delta M/M\equiv\bigl(M-M_{\rm bm}\bigr)/M$—to quantify geometry–luminous decoupling across sequences, and we confront these trends with multimessenger constraints by keeping $R$ and $R_{\rm bm}$ explicitly separate. Our central new result is that a \emph{per cent level} vacuum–like admixture is sufficient to \emph{quench} halo formation even in otherwise halo–friendly regimes, driving $\Delta R$ from multikilometer to subkilometer scales while leaving $\Delta M/M$ at the subpercent level.

The paper is organized as follows. Section~\ref{sec:structure} formulates the two–fluid stellar–structure equations with energy exchange, specifies the baryonic EOS and the fermionic DM and vacuum–like DE components, and defines the central–partition parameters used to generate sequences. Section~\ref{sec:results} presents the numerical results, highlighting halo diagnostics on $\Delta R$ and $\Delta M/M$, the luminous–versus–gravitational radius comparison on the $M$–$R$ plane, and the impact of a percent–level vacuum–like fraction. Section~\ref{sec:conclusions} summarizes the main findings, outlines multimessenger implications, and delineates modeling limitations and observational strategies for near-term tests.

\section{Dark influence on the structure equations}
\label{sec:structure}

We consider static, spherically symmetric NSs composed of ordinary (baryonic) matter and a unified dark sector, which includes DM and DE. 
The spacetime geometry is described by
\begin{equation}
  ds^2 = e^{2\nu(r)}\,dt^2 - \frac{dr^2}{1-2m(r)/r} - r^2\!\left(d\theta^2+\sin^2\!\theta\,d\phi^2\right),
  \label{metricsch}
\end{equation}
where $m(r)$ is the enclosed gravitational mass. Each sector $i\in\{\mathrm{bm},\chi,\mathrm{de}\}$ is modeled as a perfect fluid with energy–momentum tensor
\begin{equation}
  T_{(i)}^{\mu\nu} = \bigl(\epsilon_i+p_i\bigr)u^\mu u^\nu - p_i g^{\mu\nu},
  \label{eq:perfectfluid}
\end{equation}
and the total tensor is the explicit sum
\begin{equation}
  T^{\mu\nu}=\sum_{i\in\{\mathrm{bm},\chi,\mathrm{de}\}} T_{(i)}^{\mu\nu}.
\end{equation}

To account for a potential non-gravitational coupling, we follow the unified Model~III of Ref.~\cite{Araujo:2025tlv} and allow the dark sector to exchange energy with baryons. This is described by a source term $Q$ such that the total system remains covariantly conserved while the individual sectors are not:
\begin{align}
  T^{\mu\nu}_{\ ;\mu} &= 0, \qquad
  T^{\mu\nu}_{(\mathrm{bm})\ ;\mu} = Q,\qquad
  T^{\mu\nu}_{(\mathrm{dark})\ ;\mu} = -\,Q,
  \label{eq:exchange_rules}
\end{align}
with $T^{\mu\nu}_{(\mathrm{dark})}\equiv T^{\mu\nu}_{(\chi)}+T^{\mu\nu}_{(\mathrm{de})}$. For the interaction term, we adopt the phenomenological form from \cite{Araujo:2025tlv},
\begin{equation}
  Q \equiv \alpha\,\frac{d\epsilon_{\mathrm{dark}}}{dr}
    = \alpha\,\frac{d\epsilon_{\mathrm{dark}}}{dp_{\mathrm{dark}}}\,\frac{dp_{\mathrm{dark}}}{dr},
  \label{eq:Q_definition_model3}
\end{equation}
where $\alpha$ is a dimensionless coupling (we set $\alpha=1$ unless stated otherwise). By this convention, $Q>0$ denotes an energy flow \emph{into} the baryons, while $Q<0$ represents energy flowing \emph{out of} them.

The physical motivation for this phenomenological coupling is the following. In the standard two–fluid treatment, each sector is conserved separately, so baryons and dark matter evolve without any non-gravitational energy exchange. More generally, however, microscopic interactions between the visible and dark sectors may induce an effective transfer in the stellar medium. Rather than adopting a specific particle-physics model, we parametrize this effect at the hydrodynamic level by taking the interaction source to be proportional to the local radial gradient of the dark-sector energy density, $d\epsilon_{\mathrm{dark}}/dr$.  This choice ensures that the exchange vanishes where the dark component is spatially uniform and becomes strongest where its profile varies most rapidly. The dimensionless parameter $\alpha$ controls the strength of this coupling: $\alpha=0$ recovers the non-interacting limit, while nonzero $\alpha$ introduces a transfer between the baryonic and dark sectors. At the same time, the total energy–momentum tensor remains covariantly conserved, so consistency with the Einstein equations is preserved. This gradient-driven ansatz was introduced in Ref.\cite{Araujo:2025tlv} (Model~III therein); here we retain the same prescription and apply it to the halo-forming mass regime.

Combining the field equations with the conservation laws in Eqs.~\eqref{metricsch}--\eqref{eq:Q_definition_model3} yields the modified system for hydrostatic structure:
\begin{flalign}
  \frac{dp_{\mathrm{bm}}}{dr}
  + \frac{\bigl(p_{\mathrm{bm}}+\epsilon_{\mathrm{bm}}\bigr)\bigl(m+4\pi r^3 p\bigr)}{r^2-2mr}
  &= Q, 
  \label{eq:dp_m_unified}\\[2pt]
  \frac{dp_{\mathrm{dark}}}{dr}
  + \frac{\bigl(p_{\chi}+\epsilon_{\chi}\bigr)\bigl(m+4\pi r^3 p\bigr)}{r^2-2mr}
  &= -\,Q, 
  \label{eq:dp_dark_unified}\\[2pt]
  \frac{dm}{dr} &= 4\pi r^2 \epsilon, 
  \label{eq:mass_unified}\\[2pt]
  \frac{d\nu}{dr} &= \frac{m+4\pi r^3 p}{r^2-2mr}.
  \label{eq:nueq_unified}
\end{flalign}
Here, $p=p_{\mathrm{bm}}+p_{\mathrm{dark}}$ and $\epsilon=\epsilon_{\mathrm{bm}}+\epsilon_{\mathrm{dark}}$ are the total pressure and energy density. Note that in the dark sector's hydrostatic equation~\eqref{eq:dp_dark_unified}, the active gravitational mass term $(\epsilon_{\mathrm{dark}}+p_{\mathrm{dark}})$ simplifies to $(\epsilon_\chi+p_\chi)$. This is because the vacuum–like component satisfies $p_{\mathrm{de}}=-\epsilon_{\mathrm{de}}$, and thus its contribution vanishes. This system is solved subject to regularity conditions at the center, namely $m(0)=0$ and finite pressures $p_i(0)$. The stellar surface $R$ is located where the \emph{total} pressure vanishes, $p(R)=0$, which defines the total mass $M \equiv m(R)$. We also define the baryonic radius, $R_{\mathrm{bm}}$, as the location where the pressure of the baryonic fluid vanishes, $p_{\mathrm{bm}}(R_{\mathrm{bm}})=0$. If the dark component's pressure extends beyond this point, becoming zero at a larger radius, the star is surrounded by a dark halo.

To close the hydrostatic system in Eqs.~\eqref{eq:dp_m_unified}–\eqref{eq:nueq_unified}, we must specify the barotropic relations $p_i(\epsilon_i)$ for each constituent. We adopt the following equations of state:

\begin{itemize}

\item Baryonic matter: We model the cold, catalyzed baryonic component as charge-neutral, $\beta$-equilibrated $npe\mu$ matter. Its properties are described by the Akmal - Pandharipande - Ravenhall (APR) EOS~\cite{Akmal:1998cf}, which is built from variational chain-summation calculations using realistic two- and three-nucleon forces (Argonne $v_{18}$ + Urbana IX) with relativistic boost corrections. For computational purposes, we employ the Generalized Piecewise Polytropic (GPP) representation from Ref.~\cite{o2020parametrized}, which approximates the microphysical APR curve with polytropic segments while ensuring continuity of pressure, energy density, and sound speed. This yields a thermodynamically consistent surrogate across all relevant densities. The outer layers are treated by matching this core EOS to a GPP fit of the unified SLy(4) crust model~\cite{Douchin:2001sv}. This implementation strategy for the baryonic EOS has been used in our recent work~\cite{Araujo:2024txe,Araujo:2025tlv}.

\item Dark matter: DM candidates span many orders of magnitude in mass and statistics, from ultralight bosons in the wave regime to multi–TeV particulate scenarios \citep{Hui:2016ltb,Marsh:2015xka}. From the standpoint of particle microphysics, a particularly well–motivated and widely used benchmark for the dark component inside NSs is fermionic DM modeled as a cold, degenerate Fermi fluid. This choice is conservative and phenomenologically robust: degeneracy pressure provides a minimal, model–agnostic source of support without invoking condensate physics or model–dependent self–interactions, and it encompasses a broad class of WIMP–like scenarios explored in the NS context \citep{Goldman:1989nd,Kouvaris:2010jy,Li:2012ii,Leung:2022wcf,Scordino:2024ehe}. We therefore adopt a single–species, spin–$1/2$ \emph{non–interacting} Fermi gas at $T=0$ as our baseline. In the non–interacting limit the EOS of the relativistic Fermi fluid is
\begin{flalign}
p_\chi &= \frac{m_\chi^4}{24\pi^2}
\Bigl\{z\,\sqrt{z^2 + 1}\,\bigl(2z^2 - 3\bigr)
+ 3 \sinh^{-1}(z)\Bigr\},
\label{pdarkmatter}
\\
\epsilon_\chi &= \frac{m_\chi^4}{8\pi^2}
\Bigl\{z\,\sqrt{z^2 + 1}\,\bigl(2z^2 + 1\bigr)
- \sinh^{-1}(z)\Bigr\},
\label{edarkmatter}
\end{flalign}
where $m_\chi$ is the fermion mass and $z\equiv k_{F\chi}/m_\chi$ is the dimensionless Fermi momentum (we use natural units $\hbar=c=1$ and a spin degeneracy $g_\chi=2$). The number density is $n_\chi=k_{F\chi}^3/(3\pi^2)$, and the chemical potential is $\mu_\chi=m_\chi\sqrt{1+z^2}$. This ideal EOS provides a transparent baseline to quantify how the dark component affects stellar structure in our two–fluid calculations.

\item Dark energy: We model the DE sector as a vacuum–like component with local EOS $p_{\mathrm{de}}=-\epsilon_{\mathrm{de}}$. This component carries a positive energy density and negative pressure that may vary with radius. In the stellar-structure equations, it contributes to the mass profile through  $4\pi r^{2}\,\epsilon_{\mathrm{de}}(r)$ and reduces effective pressure support in hydrostatic balance, thereby lowering the maximum mass \cite{Araujo:2024txe}.

\end{itemize}

To close the system, we must provide a prescription for setting the central boundary conditions that determine the relative abundances of each component. We achieve this by parameterizing the central energy densities. First, we fix the total central energy density, $\epsilon_c \equiv \epsilon(r=0)$. Then, we introduce two dimensionless parameters, $y_{\rm bm}$ and $x_\chi$, which govern the partition of this energy density.

The parameter $y_{\rm bm}\in[0,1]$ sets the fraction of the total central density corresponding to baryonic matter:
\begin{align}
 \epsilon_{\mathrm{bm}}(0) &= y_{\rm bm}\,\epsilon_c, \label{eq:central_baryon} \\
 \epsilon_{\mathrm{dark}}(0) &= \bigl(1-y_{\rm bm}\bigr)\,\epsilon_c. \label{eq:central_dark}
\end{align}
The limiting case $y_{\rm bm}=1$ yields a purely baryonic center, whereas $y_{\rm bm}=0$ describes a center composed entirely of dark components. Equivalently, we define the central dark–sector fraction as $y_{\mathrm{dark}}\equiv 1-y_{\rm bm}$.

Next, the parameter $x_\chi\in[0,1]$ specifies the internal composition of the dark sector at $r=0$:
\begin{align}
 \epsilon_{\chi}(0) &= x_\chi\,\epsilon_{\mathrm{dark}}(0), \label{eq:central_dm} \\
 \epsilon_{\mathrm{de}}(0) &= \bigl(1-x_\chi\bigr)\,\epsilon_{\mathrm{dark}}(0). \label{eq:central_de}
\end{align}
Here, $x_\chi=1$ signifies a dark sector made entirely of dark matter, whereas $x_\chi=0$ corresponds to a pure dark-energy component.

With the central conditions fully specified by the set $\{\epsilon_c, y_{\rm bm}, x_\chi\}$, we numerically integrate the structure equations~\eqref{eq:dp_m_unified}--\eqref{eq:mass_unified} from the center ($r=0$) to the stellar surface ($r=R$), defined by the vanishing of the total pressure, $p(R)=0$. This procedure allows us to construct the mass–radius relations and investigate the impact of the dark sector's composition and interactions on the global properties of the NS.

\section{Results}
\label{sec:results}

\subsection{Halo diagnostics from radius and mass excess}

\begin{figure}[tb]
\centering
\includegraphics[width=0.95\textwidth]{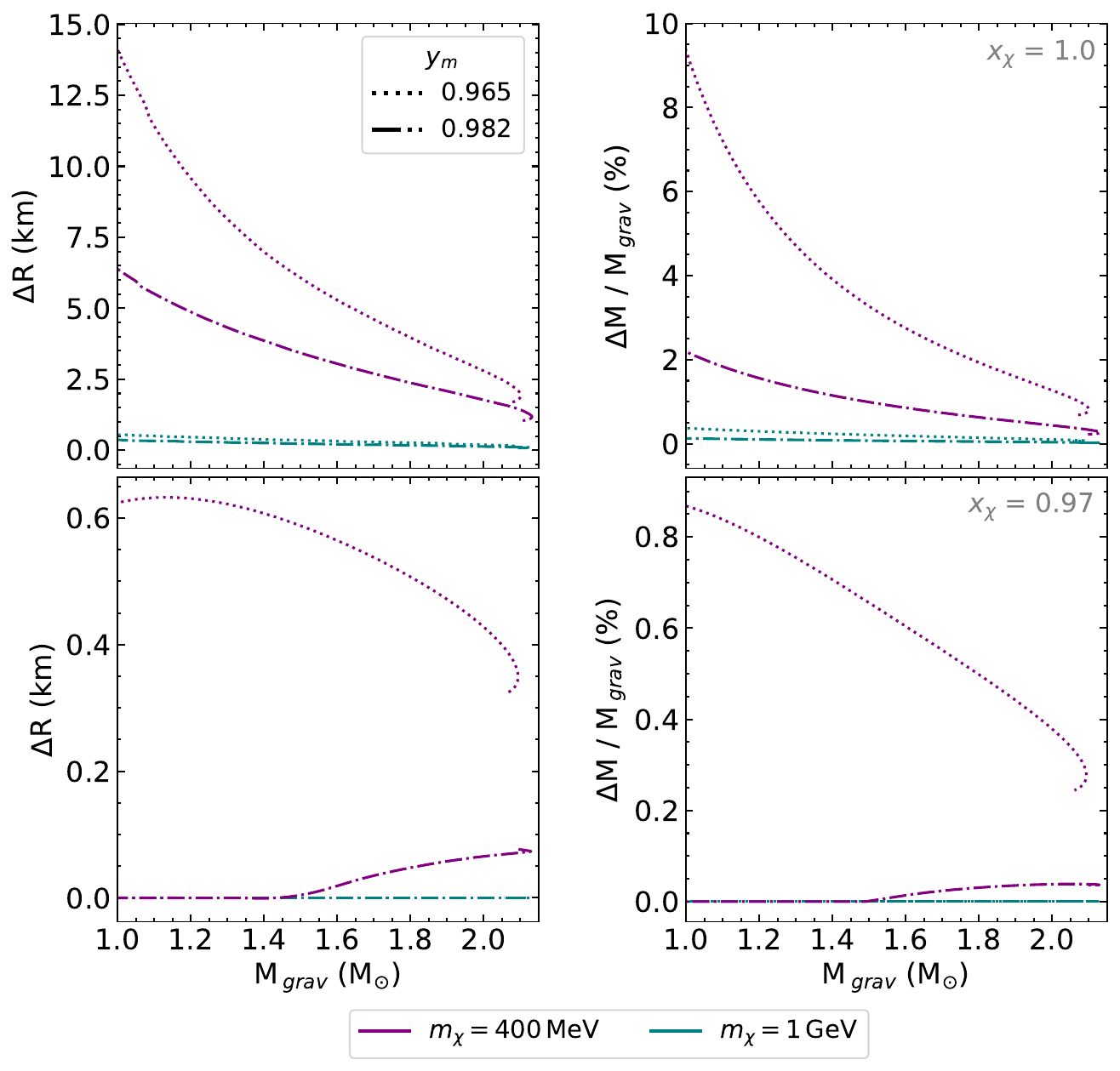}
\caption{Radius excess $\Delta R\equiv R-R_{\rm bm}$ (top) and fractional mass excess $\Delta M/M\equiv (M-M_{\rm bm})/M$ (bottom) as functions of the gravitational mass $M$ for dark–matter–admixed NSs with an interacting dark sector. Here $R$ and $M$ are the total (gravitational) radius and mass that source the exterior spacetime, while $R_{\rm bm}$ and $M_{\rm bm}$ are the baryonic (electromagnetic) counterparts inferred from the luminous component. Curves compare fermionic dark–matter masses $m_\chi=1~\mathrm{GeV}$ and $400~\mathrm{MeV}$, two central baryon fractions $y_{\rm bm}=\{0.982,\,0.965\}$, and internal dark–sector partitions $x_\chi=\{1.0,\,0.97\}$ (the latter corresponding to a 3\% vacuum–like fraction when $x_\chi=0.97$). Lighter $m_\chi$ and smaller $y_{\rm bm}$ produce extended halos that substantially increase $R$, while the associated change in $M$ is present but comparatively modest; introducing a small vacuum–like admixture ($x_\chi<1$) leads to a marked reduction in the halo’s radial extent and mass.}
\label{fig:1}
\end{figure}

Figure~\ref{fig:1} quantifies the structural impact of the dark sector by plotting two key diagnostics against the total gravitational mass $M$: the absolute radius excess, $\Delta R \equiv R - R_{\rm bm}$, and the fractional mass excess, $\Delta M/M \equiv (M - M_{\rm bm})/M$. Here, $R_{\rm bm}$ and $M_{\rm bm}$ represent the radius and mass of the baryonic component, which are in principle accessible via electromagnetic observations, whereas $R$ and $M$ are the total values that determine the exterior spacetime and gravitational response. The figure compares sequences for two fermionic dark matter masses ($m_\chi = 400~\text{MeV}$ and $1~\text{GeV}$), two central baryonic fractions ($1-y_{\rm dark}$, corresponding to $y_{\rm bm} = 0.982$ and $0.965$), and two central dark-sector compositions: a pure dark matter fluid ($x_\chi = 1.0$) and a unified fluid with a 3\% vacuum-energy component ($x_\chi = 0.97$). Three distinct physical trends emerge from these comparisons.

\paragraph{Dependence on Dark Matter Particle Mass.}
The most prominent trend is the strong dependence of halo formation on the DM particle mass, $m_\chi$. For a fixed central composition, configurations with lighter DM particles ($m_\chi = 400~\text{MeV}$) develop substantially more extended halos (larger $\Delta R$) and greater mass excesses ($\Delta M/M$) than their counterparts with heavier particles ($m_\chi = 1~\text{GeV}$). This behavior follows directly from the cold, degenerate Fermi–gas scaling: at fixed number density $n_\chi$, the pressure obeys $p_\chi \propto n_\chi^{5/3} m_\chi^{-1}$ in the nonrelativistic regime, so lighter fermions provide more pressure per unit energy density and can support a diffuse, pressure–held envelope beyond the baryonic surface, inflating $R$ while only mildly increasing $M$.

This mass ordering corroborates the canonical picture established in previous work. Two–fluid studies consistently show that sub–GeV fermions favor extended halos and large radius shifts at fixed mass, whereas $\mathcal{O}(\text{GeV})$ fermions tend to suppress halo formation —driving $R\!\to\!R_{\rm bm}$ and yielding small $R\!-\!R_{\rm bm}$—rather than producing robust dark cores; core–like configurations arise only in specific setups and/or at sufficiently large dark fractions (see, e.g., \cite{Scordino:2024ehe}), while analyses focused on light/admixed DM typically find halo–dominated or halo–absent regimes, not stable DM cores \cite{Leung:2022wcf,Nelson:2018xtr,Karkevandi:2021ygv}. At the opposite extreme, very heavy ($\gtrsim$GeV) DM components soften the effective EoS without forming halos, lowering $M_{\max}$ and potentially conflicting with the $2\,M_\odot$ bound—consistent with early effective–fluid treatments \cite{Li:2012ii}. The observational ramifications of halo formation are likewise well documented: tidal–response calculations identify large changes in Love numbers and deformabilities when a low–density dark envelope is present \cite{Leung:2022wcf,Nelson:2018xtr}, and pulse–profile (self–lensing) studies show that the footprint scales with the “halo compactness” rather than with halo mass alone \cite{Liu:2024swd,Miao:2022rqj,Shawqi:2024jmk}. These precedents motivate the mass scan adopted here: $m_\chi=400~\text{MeV}$ is adopted as a representative halo–forming benchmark and the most stringent case for testing the quenching mechanism, while $m_\chi=1~\text{GeV}$ provides a baseline where halos are already strongly suppressed. 
The corresponding halo signatures therefore do not follow a simple linear trend with $m_\chi$, but instead become markedly weaker as the DM particle mass increases \citep{Routaray:2023spb,Grippa:2024ach}.

\paragraph{Dependence on Dark Matter Abundance.}
The overall scale of the dark halo is, as expected, controlled by the dark abundance, here parameterized by the central dark fraction $y_{\rm dark}$ (equivalently, by the global mass fraction $f_\chi\!\equiv\!M_{\chi}/M$ along a sequence). Decreasing the central baryonic content (i.e., lowering $y_{\rm bm}$ from $0.982$ to $0.965$) systematically increases both $\Delta R$ and $\Delta M/M$ for all models. A larger reservoir of DM allows a greater portion of the dark fluid to settle into low–density hydrostatic equilibrium outside the luminous surface, so the gravitational radius $R$ grows while the stellar mass $M$ shifts more modestly. This \emph{radius–dominant} response with increasing $f_\chi$ is the same qualitative behavior reported in two–fluid calculations: for fixed $m_\chi$, raising the dark fraction enlarges the $R$–$R_{\rm bm}$ split and enhances tidal effects when halos form \citep{Leung:2022wcf,Nelson:2018xtr}. Quantitatively, Ref. \cite{Scordino:2024ehe} mapped the $(m_\chi,f_\chi)$ plane with a realistic nuclear EoS and identified a \emph{critical} curve $f_\chi^{\rm crit}(m_\chi)$ from the $2\,M_\odot$ condition, showing that (i) sub–GeV masses tolerate sizable $f_\chi$ (halo regimes) while (ii) $\mathcal{O}(\mathrm{GeV})$ masses require small $f_\chi$ to avoid suppressing $M_{\max}$ below $2\,M_\odot$ (halo–quenched regimes). In bosonic–DM studies with repulsive self–interactions, the same monotonicity with abundance appears—larger fractions amplify radius shifts and tidal imprints up to observational bounds on $\Lambda_{1.4}$—thereby restricting $f_\chi$ to percent–level values \citep{Karkevandi:2021ygv}. Early effective–fluid analyses already noted that increasing the dark admixture softens the global EoS and lowers $M_{\max}$, tightening the viable range of $f_\chi$ when $m_\chi$ is heavy \citep{Li:2012ii}. On the observational side, both tidal–deformability studies and pulse–profile/self–lensing analyses emphasize that the \emph{detectability} of a given $f_\chi$ depends on how much of it resides in the outer, low–density layer: halo–dominated configurations at moderate $f_\chi$ can produce large changes in $k_2$ and $\Lambda$ \citep{Leung:2022wcf,Nelson:2018xtr} and measurable flux–peak modulations set by halo compactness rather than by $f_\chi$ alone \citep{Liu:2024swd,Miao:2022rqj,Shawqi:2024jmk}. Our trends for $m_\chi=400~\mathrm{MeV}$ (large $\Delta R$ at lower $y_{\rm bm}$) and for $m_\chi=1~\mathrm{GeV}$ (halo suppression even as $y_{\rm bm}$ decreases) therefore reproduce the established abundance systematics and provide a controlled baseline to assess how adding a vacuum–like fraction further quenches the halo at fixed $f_\chi$.

\paragraph{Effect of a Dark Energy Admixture.}
A central novelty of this work is the explicit inclusion of a vacuum–like component in the unified dark sector and the demonstration—\emph{in the halo–forming mass regime}—that even a \(\sim\!3\%\) dark–energy (DE) fraction \((x_\chi=0.97)\) efficiently \emph{quenches halo formation}. As seen in Figure~\ref{fig:1}, the DE admixture drives a sharp contraction of the exterior dark layer: for the same central partition and fermion mass, the radius excess decreases from \(\mathcal{O}(10)\) km (pure DM) to sub–kilometer levels, while the fractional mass excess is suppressed from several percent (up to \(\sim 10\%\)) to the sub–percent regime—i.e., order–of–magnitude reductions in \(\Delta M/M\) and factors \(\gtrsim 10\!-\!20\) in \(\Delta R\), with the effect most pronounced for \(m_\chi=400~\mathrm{MeV}\) and still visible at \(1~\mathrm{GeV}\). Physically, this behavior follows directly from the structure equations in our Model~III: the vacuum sector contributes negative pressure but \emph{no} inertial term in the TOV force balance, since \(\epsilon_{\rm de}+p_{\rm de}=0\); as a result, at fixed \(\epsilon_{\rm dark}\) the supporting pressure comes almost solely from the fermionic DM, which is now a smaller fraction of the dark sector. The net effect is a steeper outward decline of the \emph{total} pressure and an earlier satisfaction of the surface condition \(p(R)=0\), which collapses the halo and drives \(R\!\to\!R_{\rm bm}\). In addition, while \(\epsilon_{\rm de}\) still feeds the mass function \(m(r)\), the lack of a corresponding pressure–support term enhances the gravity–to–support imbalance in the outer layers, reinforcing the halo quenching.

This DE–induced suppression is \emph{qualitatively new} relative to previous dark matter admixed NSs (DANS) studies—which either neglected DE entirely and focused on pure DM (fermionic or bosonic) \citep{Leung:2022wcf,Nelson:2018xtr,Karkevandi:2021ygv,Scordino:2024ehe,Shawqi:2024jmk,Miao:2022rqj} or, in the case of our previous work \cite{Araujo:2025tlv}, explored DE at \(m_\chi\!=\!10~\mathrm{GeV}\) where halos do not form. Those works established that sub–GeV fermions favor extended halos and large radius shifts at fixed mass, whereas \(\mathcal{O}(\mathrm{GeV})\) masses tend to \emph{suppress} halos; our results extend that picture by showing that a \emph{percent–level} DE fraction can suppress halos \emph{even within} an otherwise halo–friendly mass range, thereby collapsing \(R-R_{\rm bm}\) while leaving \(M\) comparatively unchanged.

\begin{table}[tb]
\centering

\begin{minipage}{\textwidth}
\centering
\caption{Radius excess $\Delta R \equiv R - R_{\rm bm}$ (km) for $m_\chi = 400~\mathrm{MeV}$, $y_{\rm bm} = 0.982$, and $\alpha = 1$, evaluated at two representative masses across two baryonic EOSs and three values of the DE parameter~$\omega$.}
\label{tab:robustness_DR}
\smallskip
\renewcommand{\arraystretch}{1.3}
\begin{tabular}{ll ccc ccc}
\hline\hline
 & & \multicolumn{3}{c}{$M = 1.4\,M_\odot$} & \multicolumn{3}{c}{$M = 2.0\,M_\odot$} \\
\cline{3-5}\cline{6-8}
EOS & $x_\chi$ & $\omega{=}{-}0.35$ & $\omega{=}{-}1$ & $\omega{=}{-}1.65$ & $\omega{=}{-}0.35$ & $\omega{=}{-}1$ & $\omega{=}{-}1.65$ \\
\hline
\multirow{2}{*}{APR}
 &1.0 & 3.76 & 3.76 & 3.76  & 1.73 & 1.73 & 1.73 \\
 &0.97& 1.23 & 0.00 & 0.00  & 0.71 & 0.06 & 0.00 \\[3pt]
\multirow{2}{*}{MPA1}
 & 1.0 & 3.84 & 3.84 & 3.84  & 2.12 & 2.12 & 2.12 \\
 & 0.97 & 1.03 & 0.00 & 0.00 & 0.71 & 0.00 & 0.00 \\
\hline\hline
\end{tabular}
\end{minipage}

\vspace{0.8em}

\begin{minipage}{\textwidth}
\centering
\caption{Fractional mass excess $\Delta M/M$ (\%) for the same configurations as in Table~\ref{tab:robustness_DR}.}
\label{tab:robustness_DM}
\smallskip
\renewcommand{\arraystretch}{1.3}
\begin{tabular}{ll ccc ccc}
\hline\hline
& & \multicolumn{3}{c}{$M = 1.4\,M_\odot$} & \multicolumn{3}{c}{$M = 2.0\,M_\odot$} \\
\cline{3-5}\cline{6-8}
EOS & $x_\chi$ & $\omega{=}{-}0.35$ & $\omega{=}{-}1$ & $\omega{=}{-}1.65$ & $\omega{=}{-}0.35$ & $\omega{=}{-}1$ & $\omega{=}{-}1.65$ \\
\hline
\multirow{2}{*}{APR}
& 1.0 & 1.11 & 1.11 & 1.11 & 0.43 & 0.43 & 0.43 \\
& 0.97 & 0.57 & 0.00 & 0.00 & 0.25 & 0.04 & 0.00 \\[3pt]
\multirow{2}{*}{MPA1}
& 1.0 & 1.06 & 1.06 & 1.06 & 0.48 & 0.48 & 0.48 \\
& 0.97 & 0.45 & 0.00 & 0.00 & 0.25 & 0.00 & 0.00 \\
\hline\hline
\end{tabular}
\end{minipage}

\end{table}

\paragraph{Robustness against the baryonic EOS and stellar mass.}
All sequences discussed above were computed using APR as our baseline hadronic EOS. To verify that the main conclusions are not an artifact of this specific choice, we repeated the halo diagnostics for a second hadronic EOS, MPA1, which is appreciably stiffer than APR. Rather than reproducing the full set of figures for both EOSs, which would add substantial redundancy without changing the physical interpretation, we summarize this robustness test in Tables~\ref{tab:robustness_DR} and \ref{tab:robustness_DM} for two representative gravitational masses, \(M=1.4\,M_\odot\) and \(M=2.0\,M_\odot\), and for the most halo-prone benchmark considered here, \(m_\chi=400~\mathrm{MeV}\). The tables show that the qualitative pattern is unchanged under both variations. Pure-DM configurations (\(x_\chi=1\)) develop kilometer-scale halos at both masses for both hadronic EOSs, whereas the inclusion of a small dark-energy--like fraction strongly suppresses both \(\Delta R\) and \(\Delta M/M\). This suppression is complete, or nearly so, in the vacuum-like and phantom-like cases, and remains substantial even for the quintessence-like benchmark \(\omega=-0.35\). Quantitatively, the suppression is in fact slightly more efficient for MPA1 than for APR in the configurations displayed, especially at \(M=1.4\,M_\odot\), although the differences are modest and do not alter the physical interpretation. The comparison between \(1.4\,M_\odot\) and \(2.0\,M_\odot\) further shows that the effect is not confined to a narrow mass interval: heavier stars generally exhibit smaller absolute values of \(\Delta R\) and \(\Delta M/M\) at fixed microphysical parameters, consistent with their deeper gravitational potentials confining the outer dark layers more efficiently. Overall, the halo-quenching mechanism reported here is therefore robust against both the choice of hadronic EOS and changes in stellar mass across the astrophysically relevant range.

\paragraph{Sensitivity to departures from the vacuum limit.}
To assess the sensitivity of the halo-quenching mechanism to the 
equation of state of the dark-energy--like component, we evaluate 
the halo diagnostics for three representative values of the DE 
parameter, $\omega=-0.35,\,-1,\,-1.65$, spanning the 
quintessence-like, vacuum-like, and phantom-like regimes. The exact 
vacuum limit $\omega=-1$ is a natural reference point because the 
inertial combination $\epsilon_{\mathrm{de}}+p_{\mathrm{de}}$ 
vanishes identically, so the DE-like component contributes to the 
gravitational mass without providing any compensating pressure 
support, leading to very efficient halo quenching. For $\omega>-1$ 
this combination becomes positive, partially restoring pressure 
support in the outer layers and allowing a residual halo to 
survive; for $\omega<-1$ it turns negative, which further enhances 
the gravity--support imbalance and makes the suppression even 
stronger than in the vacuum case. Quantitatively, the results in 
Tables~\ref{tab:robustness_DR} and \ref{tab:robustness_DM} show 
that for $\omega=-0.35$ the radius and mass excesses are still 
substantially reduced relative to the pure-DM case, but are not 
completely eliminated at the benchmark partition $x_\chi=0.97$. By 
contrast, for the same dark-sector partition, the halo is 
effectively quenched for $\omega\leq -1$, with $\Delta R$ and 
$\Delta M/M$ vanishing or falling below the quoted precision in 
most configurations. We have verified that the incomplete 
suppression found for the quintessence-like benchmark is not a 
qualitative obstruction: increasing the dark-energy--like fraction 
only modestly---to about $7\%$ for APR ($x_\chi\simeq 0.93$) and 
$\sim 6.5\%$ for MPA1 ($x_\chi\simeq 0.935$)---is already 
sufficient to drive the halo diagnostics to an effectively 
vanishing level. These results demonstrate that the suppression 
mechanism is robust against departures from the exact vacuum limit, 
although its efficiency is modulated by the value of $\omega$: for 
$\omega<-1$ the quenching is at least as strong as in the vacuum 
case, while for $\omega>-1$ a somewhat larger DE-like fraction is 
required to achieve complete suppression. This $\omega$-scan is 
also timely in view of the renewed cosmological interest in 
possible departures from $w=-1$, although any direct connection 
between a cosmological $w(z)$ and the local barotropic parameter 
used here lies beyond the scope of the present work.

\subsection{Mass–radius relations with luminous vs gravitational radii}

\begin{figure}[tb]
\centering
\includegraphics[width=0.95\textwidth]{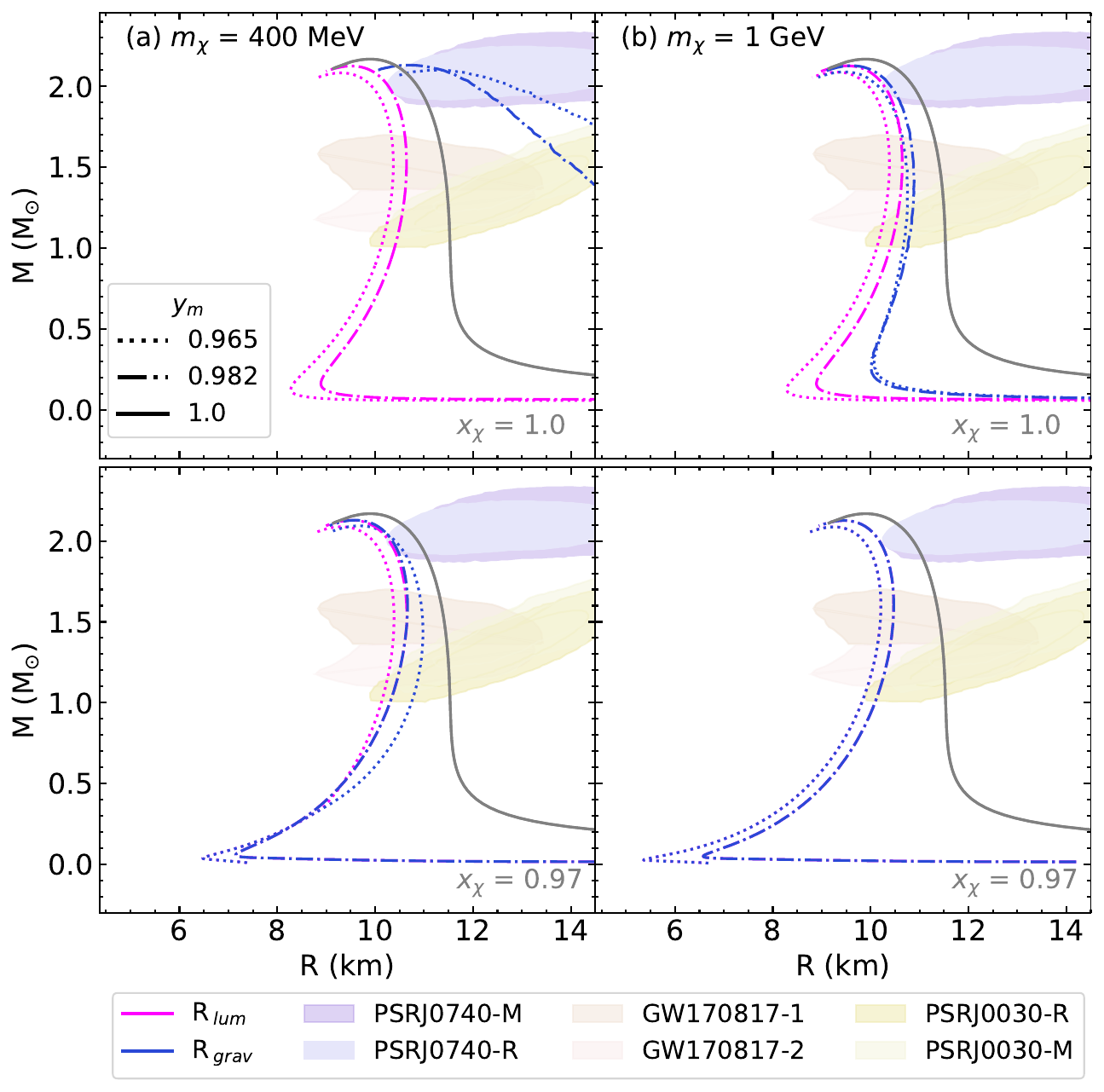}
\caption{Mass–radius relations of NSs modified by dark sector interactions, constrained by observations (background clouds). Two primary markers are presented: pink curves, representing the contribution from the luminous radius alone, and blue curves, indicating the total gravitational radius of the objects. The solid gray curve denotes the baseline case in which only baryonic matter is considered ($y_m = 1.0$).}
\label{fig:2}
\end{figure}

Figure~\ref{fig:2} recasts the diagnostics of Figure~\ref{fig:1} in the $M$–$R$ plane, showing, for each sequence, both the baryonic radius $R_{\rm bm}$ and the total radius $R$. The microphysical setups are identical to those in Figure~\ref{fig:1}—$m_\chi=400~\mathrm{MeV}$ and $1~\mathrm{GeV}$, $x_\chi=1.0$ and $0.97$, and $y_{\rm bm}\in\{0.982,\,0.965\}$—with the sole addition of a purely baryonic baseline, $y_{\rm bm}=1$, used here as an on-plot benchmark. In this representation, the radius excess $\Delta R\equiv R-R_{\rm bm}$ appears as a horizontal offset between paired curves: whenever a low-density dark envelope forms, $R$ lies to the right of $R_{\rm bm}$. For $m_\chi=400~\mathrm{MeV}$, extended regions with $R$ significantly larger than $R_{\rm bm}$ emerge, whereas for $m_\chi=1~\mathrm{GeV}$ the differences are smaller but still appreciable (upper panels). Introducing a percent-level vacuum–like fraction ($x_\chi=0.97$) largely collapses that separation and drives $R\!\to\!R_{\rm bm}$ (lower panels). It is evident that the radius response dominates over the mass response: the locus of maximum mass and the overall $M$ scale shift only modestly compared with the horizontal displacement in $R$ as the dark-sector parameters vary.

We overlay NICER inferences (PSR J0030+0451 and PSR J0740+6620) and GW constraints from GW170817 (LIGO/Virgo) \cite{Abbott:2018exr, Miller:2021qha, Miller:2019cac, Riley:2021pdl, Riley:2019yda}. Since $\Delta M/M$ is typically sub-percent in the regimes probed here, we display the total gravitational mass $M$ for all tracks. Crucially, the two classes of measurements constrain different radii: NICER hotspot modeling refers to the photospheric surface and should be compared against the $M$–$R_{\rm bm}$ tracks, whereas tidal deformability from binary inspirals probes the exterior spacetime and thus constrains the $M$–$R$ tracks. 
Observationally, the GW170817 credible region lies slightly to the left (smaller $R$) of the NICER posteriors—a model-independent ordering. By contrast, halo formation generically predicts $R>R_{\rm bm}$ at fixed $M$, which would place the GW-traced $M$–$R$ locus to the right of the NICER-traced $M$–$R_{\rm bm}$ locus. The fact that we observe the opposite implies that any halo, if present, must be thin, with $\Delta R\!\lesssim\!2$–$3~\mathrm{km}$—a scale set by the width of the radial interval where the NICER posteriors and the GW170817 credible region overlap in the $M$–$R$ plane (i.e., the common radius band permitted by both). Such suppression arises naturally if the dark-matter particle mass lies near the upper end of the sub-GeV range and/or if a percent-level vacuum–like fraction is present, which reduces $\Delta R$.

\section{Summary and conclusions}
\label{sec:conclusions}

We have investigated NSs hosting a unified dark sector composed of cold, degenerate fermionic DM and a vacuum–like DE component. Our analysis employed a general-relativistic two-fluid hydrostatic framework that allows for a covariantly conserved, gradient–driven energy exchange between baryons and the dark sector. The study focuses on the \emph{halo–forming} mass range for fermionic DM, $m_\chi=\{400~\mathrm{MeV},\,1~\mathrm{GeV}\}$, and on observables that disentangle luminous from gravitational structure, namely the radius and mass excesses $\Delta R \equiv R-R_{\rm bm}$ and $\Delta M/M \equiv (M-M_{\rm bm})/M$.

A first outcome of this study is the confirmation of well–established systematics regarding the roles of the DM particle mass and its abundance. Using the same diagnostics across sequences, we find that sub–GeV fermions (e.g., $m_\chi=400~\mathrm{MeV}$) give rise to extended, low–density halos that increase the gravitational radius $R$ by several kilometers, while the accompanying change in mass remains comparatively modest. By contrast, near–GeV masses (e.g., $m_\chi=1~\mathrm{GeV}$) markedly reduce halo extent and bring $R$ closer to $R_{\rm bm}$ at comparable $M$. The abundance trend is likewise clear: for fixed $m_\chi$, decreasing the central baryon fraction $y_{\rm bm}$ amplifies both the radius and mass excesses. These behaviors are consistent with two–fluid calculations of fermionic DANS and with related bosonic–DM analyses~\citep{Leung:2022wcf,Nelson:2018xtr,Scordino:2024ehe,Karkevandi:2021ygv,Li:2012ii}.

A central result of this study concerns the effect of a percent–level vacuumlike fraction in the unified dark sector. In our sequences, introducing a small DE admixture (e.g., $x_\chi=0.97$, corresponding to a 3\% DE fraction) markedly reduces halo extent and mass even in otherwise halo–friendly regimes: $\Delta R$ drops from several kilometers to sub–kilometer scales, while $\Delta M/M$ is suppressed from a few percent to the sub–percent level. The effect is most pronounced for $m_\chi=400~\mathrm{MeV}$ and remains visible at $1~\mathrm{GeV}$. The mechanism is hydrostatic and follows directly from the structure equations: the vacuum component contributes negative pressure to the TOV balance, so the total pressure support in the outer layers decreases while $m(r)$ still receives a contribution from $\epsilon_{\rm de}$.  This enhances the outward decline of the total pressure, reducing the radius at which $p(R)=0$ and driving $R$ closer to $R_{\rm bm}$, with a comparatively smaller impact on $M$. To our knowledge, no prior DANS analysis has reported and quantitatively characterized percent–level DE–induced halo contraction in the halo–forming regime.

These results have immediate multimessenger implications. Because gravitational waves probe $R$, while X–ray pulse–profile modeling constrains a photospheric radius tied to $R_{\rm bm}$, halo–forming configurations naturally predict a split between the radii inferred by the different channels. In fact, halo–bearing models generically predict $R>R_{\rm bm}$ at fixed $M$, which would place the GW–inferred $M$–$R$ locus to the \emph{right} of the X-ray–inferred $M$–$R_{\rm bm}$ locus. Current data show the opposite ordering: the credible region from GW170817 lies slightly to the \emph{left} (smaller radii) of the NICER posteriors, strongly disfavoring large halos and implying a tight bound $\Delta R\equiv R-R_{\rm bm}\lesssim 2$–$3~\mathrm{km}$. Within our framework this small offset is naturally achieved not only for fermionic DM with masses near the GeV scale, but also for lighter DM provided a percent-level vacuum–like fraction is present, both cases shrinking the dark envelope while leaving $M$ comparatively less affected.

Looking ahead, these conclusions suggest a clear observational strategy for future multimessenger surveys. Improved gravitational-wave measurements of tidal deformability, and eventually of post-merger dynamics, will tighten constraints on the total gravitational radius $R$, while next-generation X-ray observations may sharpen the inference of the luminous radius $R_{\rm bm}$. In this context, the key target is not only the possible detection of a mismatch between the two radii, but also the ability to place increasingly stringent upper bounds on $\Delta R$ when no such mismatch is observed. Within our framework, such bounds translate directly into constraints on both the extent of any dark halo and the allowed vacuum–like fraction of the dark sector, making future gravitational-wave facilities and high-precision X-ray missions particularly relevant for testing this scenario.





\acknowledgments

LFA. was supported by Coordenação de Aperfeiçoamento de Pessoal de Nível Superior (CAPES), Brazil. GL acknowledges partial financial support from the Brazilian agency CNPq (Grant No. 316844/2021-7). JASL is partially supported by the Brazilian agencies, CNPq under Grant No. 310038/2019-7, CAPES (88881.068485/2014) and FAPESP (LLAMA Project No. 11/51676-9).



\bibliographystyle{JHEP}
\bibliography{biblio.bib}


\end{document}